\documentclass[submission,copyright]{eptcs}
\providecommand{\event}{MBT 2012} % Name of the event you are submitting to
\usepackage{graphicx}
\usepackage{listings}
\title{Using Built-In Domain-Specific Modeling Support to Guide Model-Based Test Generation}
\author{Teemu Kanstr\'{e}n
\institute{VTT\\ Oulu, Finland}
\email{teemu.kanstren@vtt.fi}
\and
Olli-Pekka Puolitaival
\institute{F-Secure\\
Oulu, Finland}
\email{olli-pekka.puolitaival@fsecure.com}
}
\def\titlerunning{Built-In DSM+MBT}
\def\authorrunning{T. Kanstr\'{e}n, O-P. Puolitaival}
\begin{document}
\maketitle

\lstset{
language=Java,
basicstyle=\small\sffamily,
numbers=left,
numberstyle=\tiny,
frame=single,
columns=fullflexible,
showstringspaces=false
}

\begin{abstract}
We present a model-based testing approach to support automated test generation with domain-specific concepts. This includes a language expert who is an expert at building test models and domain experts who are experts in the domain of the system under test. First, we provide a framework to support the language expert in building test models using a full (Java) programming language with the help of simple but powerful modeling elements of the framework. Second, based on the model built with this framework, the toolset automatically forms a domain-specific modeling language that can be used to further constrain and guide test generation from these models by a domain expert. This makes it possible to generate a large set of test cases covering the full model, chosen (constrained) parts of the model, or manually define specific test cases on top of the model while using concepts familiar to the domain experts.
\end{abstract}

\section{Introduction}

Model-based testing (MBT) as an advanced test automation concept has been gaining increasing interest in the industry in recent years. MBT can be defined in various ways, and in this paper we follow the definition by Utting and Legeard \cite{utting2007} as "Generation of test cases with oracles from a behavioral model". MBT can be a powerful approach in generating test cases to cover various aspects of the system under test from the test models. However, typically several different types of expertise are required to build useful test models and to effectively generate test cases from these models.

Firstly, creating models for MBT is essentially a programming activity, describing the system under test (SUT) at a high level and from the testing viewpoint. Although graphical modeling notations exist, such as using UML state charts, at some level the user always needs to describe the behavioral aspects in terms of some form of programming language constructs \cite{utting2007}. This aspect is highlighted by the use of the term model program to describe the test models \cite{grieskamp2011}.

Thus we can say that programming skills are required to produce the test models. Additionally, each MBT tool has its own notation for describing the test aspects (e.g., defining what constitutes a test step) on top of this test model \cite{utting2011}, requiring the user to also have expertise in the specific modeling notations of the used tools. However, having programming skills and knowing the tool notations is typically not enough but one also requires domain expertise in order to be able to build useful test models. Typically all the required expertise can only be found in combination of several experts.

In this paper we view MBT from the viewpoint of domain-specific modeling (DSM) and how we can integrate support for more effective modeling and test generation for MBT tools with the help of DSM concepts. In DSM terminology, the person with the programming and tool expertise is termed here as the language expert and the person with domain knowledge as the domain expert.

We discuss the use of a MBT tool that transforms the test model into a DSM language, and how this can be used to more effectively guide test generation. We show how the language developer can use our OSMOTester MBT tool framework to develop test models using a full (Java) programming language and all the benefits it provides (reuse of skills, test libraries, IDE integration with debugging, refactoring and more). It is our experience that this type of a modeling language provides a powerful modeling basis and is as easier if not easier to learn than custom semi-graphical notations of many MBT tools. The OSMOTester toolset provides a framework for creating these test models, and from these test models automatically creates a DSM language that the domain expert can use to guide test generation.

The rest of this paper is structured as follows. Section~\ref{sec:background} presents the background concepts of model-based testing and domain-specific modeling in more detail, and also discusses related work. Section~\ref{sec:osmotester} presents the OSMOTester toolset and the approach it supports in detail. Section~\ref{sec:discussion} provides discussion on the topic. Finally, section~\ref{sec:conclusions} concludes the paper.

\section{Background}
\label{sec:background}

This section briefly presents the relevant background concepts for this paper and discusses related work.

\subsection{Model-Based Testing}

As described before, model-based testing as discussed here is about using behavioral models as a basis to generate tests for a system. This requires various tools and expertise to provide a useful result. Several different MBT tools exist, each with their own set of features and test generation algorithms \cite{utting2011}. These also use various notations that are suited for generic representation of system behavior, such as state-based, transition-based, and function-based notations, and various combinations of such notations \cite{utting2011}. As noted, practically all of them require representing the system under test in terms of some programming language constructs to enable test generation with input data and oracles from the model. 

In this regard, we borrow the term model program from \cite{grieskamp2011} to describe the test models. Grieskamp et al. \cite{grieskamp2011} define the model program as using guarded-update rules to modify the global data state. When a rule is invoked, a transition between the data states takes place and at the same time a method on the SUT test adapter is invoked. As a result, test steps (sometimes called actions) take place in an order defined by the model, the model traversal algorithms, and as executed by the test adapter.

%With regards to the required expertise, Figure \ref{fg:model-elements} shows an overview of the different expertise required in providing test models for MBT.
With regards to the required expertise, it can be said that several different forms of expertise are required to produce useful models.
Domain knowledge is required to produce test cases that link with the SUT, produce meaningful input data and evaluate the results in a useful way. Modeling expertise is required to understand how different aspects of the SUT behavior can be effectively described in the test models. Tool knowledge is required to understand the notation of the MBT tools used and to build models that effectively make use of the provided features of the tools modeling language and allow the tool to effectively generate test cases from these models. Test expertise in general is required to understand not only the vertical target domain but also the horizontal domain of test automation and how they should map together to produce a useful overall test environment.

%\begin{figure}
%\begin{center}
%\includegraphics[height=2.05in]{modelling-elements.pdf} 
%\caption{Modeling elements}
%\label{fg:model-elements}
%\end{center}
%\end{figure}

In a practical context, with complexity of modern systems, it is unrealistic to expect one or few persons to have all this knowledge of different domains. The target domain expert is usually an expert on the application domain and should not be expected to know all the details of MBT to produce useful test cases. A test expert in the target domain may be an expert in applying test automation concepts in the specific domain with the help of the domain expert. Another expert may be an expert in using specific MBT tools and in expressing system behavior in terms of behavioral models.

Different combinations of experts typically exist for different target domains and systems, some cross-domain experts in several areas, and several experts may be available in any domain. However, effectively working together is required by the different experts to produce useful test models that in the end allow for effective test generation for the target system. With sufficient resources, the overall MBT process can be carried out using only MBT tools and models, for example, using purely model programs as test models, with collaboration of the different (language and domain) experts at all points of the process. However, it is also useful to provide support for the domain experts to perform independent exploration of the test target with the help of the test models. To support this, we look at the concepts of domain-specific modeling.

\subsection{Domain-Specific Modeling}

Domain-specific modeling can be defined as using models to raise the level of abstraction, using domain concepts to describe the solution \cite{kelly2008}. From these models, it is possible to automatically generate the required product (code) as both the modeling language and the generators are created specifically for the company and the domain (application). The DSM approach consists of two main parts:

\begin{itemize}\addtolength{\itemsep}{-0.5\baselineskip}

\item Language and generator development, and

\item Solution modeling.

\end{itemize}

%\begin{figure}
%\begin{center}
%\includegraphics[height=3.05in]{process.png} 
%\caption{The Modeling Process}
%\label{fg:process}
%\end{center}
%\end{figure}

In the first part, the modeling notation and the transformation from that notation to the actual end product are defined. The first part involves the domain expert and the language developer. The second part involves the developers who work on developing the actual end products. In case of DSM, this also includes domain experts as they can use the higher abstraction level of the domain specific language to develop products as well.

DSM works best and is most cost-effective when developing several end products which share characteristics but also vary in different ways. In this case the same modeling language can be used to create several products, justifying the cost of language development. In traditional domains the target systems have typically been product families, in which different products can be modeled using the same domain specific language. In the test automation domain, we observe that even a single system typically has numerous varying test cases, based on the same base language. 

In our previous work we have used DSM to express test cases directly \cite{op2011} and also to express test models in MBT with a transformation to a specific chosen MBT tool \cite{teemu2012}. This means the domain expert is able to build test cases and test models using familiar domain concepts. However, these approaches require a language expert to define the language in a DSM tool which has no built-in support for the MBT tool notation, making the language development challenging without advanced editing support, requiring specific DSM tool skills, making development harder and complicating maintenance.

In this paper we describe a modeling approach for MBT where the DSM language creation is integrated into the model built using the MBT tool with only little or no extra effort required from the language developer. As the model is built on an existing popular programming language, it allows for re-use of skills, development environments, and other assets.

\subsection{Related works}

Model-based testing has been reported as having been successfully applied in several domains. This includes aerospace \cite{blackburn2002}, automotive \cite{bringmann2008,pretschner2005}, medical \cite{vieira2008}, communication protocols \cite{grieskamp2011}, and information systems \cite{neto2008}. These studies typically show improvements and benefits in applying the MBT approach in the different domains. However, they also highlight the typical situation where the model is built by an expert in the MBT tools and their notation, with the help of information provided by the domain experts. Providing means for a domain expert without the need of deep MBT tool expertise is a topic where domain-specific modeling can be applied.

In MBT, the test model is typically created to describe a large set of potential SUT behavior. The number of potential test combinations that can be generated from such models is huge or even unbounded. To address this issue, Grieskamp et al. \cite{grieskamp2011} have applied model slicing, where the expert defines constraints in the notation of the model program on how the test generator will generate test cases from the model. They use the term scenario-based test generation viewing the slices to define certain test generation scenarios. Our approach is similar while also providing a higher-abstraction level DSM language for the domain expert. Additionally, we also provide means to use this to manually create specific test scripts from the test models.

Specialized approaches for using MBT in specific domains have also been proposed. Takala et al. \cite{takala2011} have built a tool for model-based testing of Android smart-phone applications. This tool is intended to build test models for Android applications in terms of their common user-interface elements. This provides a specific DSM for MBT in for the Android application domain. We provide a generic approach that can be used as a basis for more specific approaches such as these.

Another example of applying DSM with MBT is presented by Kloos and Eschbach for the domain of railway control systems \cite{kloos2010}. A specific language is presented for this domain and this can be used to create test models. This language requires detailed knowledge on domain aspects and formalisms such as Mealy machines and process calculi, which can be challenging for domain experts. We use a common programming language (Java) as the underlying notation to ease the language development, and provide a domain-specific abstraction on top of it to enable the domain expert to work with the model as well.

A related approach is presented by Katara and Kervinen \cite{katara2006} who use low-level keywords (e.g., press key X) and on top of those, higher-level domain-specific action words (e.g., take picture), and transitions between them, to describe the test model for MBT. Our approach is similar in using textual domain-specific language in the context of MBT. However, we automatically produce these domain languages from the test model and allow their use to guide test generation and effectively generate specialized versions of the model.

Finally, our previous work on creating graphical modelling DSM languages for MBT is also relevant \cite{teemu2012}. In the previous work specific modeling languages are built for specific domains and complete model programs are generated, which are challenging to create, maintain and evolve in a graphical code generator not built for the MBT tools notation. In this paper, we present an approach where the DSM is automatically built as part of the provided test model. An interesting extension would be to add a graphical DSM layer on top of our automatically generated textual DSM language.

\section{The OSMOTester Approach}
\label{sec:osmotester}

Our approach is implemented and available in a tool called OSMOTester \cite{osmotester}. Here we refer to our approach as the OSMOTester approach according to this implementation. This approach is based on two layers in line with the definition of domain specific modeling we presented in section \ref{sec:background}: one for language and generator (model program) development, and another for the solution development (DSM guided test generation). In this section we first present a high-level overview of the approach, followed by the support for developing the language and the test generator, and finally the support for the solution development. This approach works both for online (direct test execution in generation) and offline (generate scripts and execute later) MBT approaches.

\subsection{High-Level Overview}

\begin{figure}
\begin{center}
\includegraphics[height=2.05in]{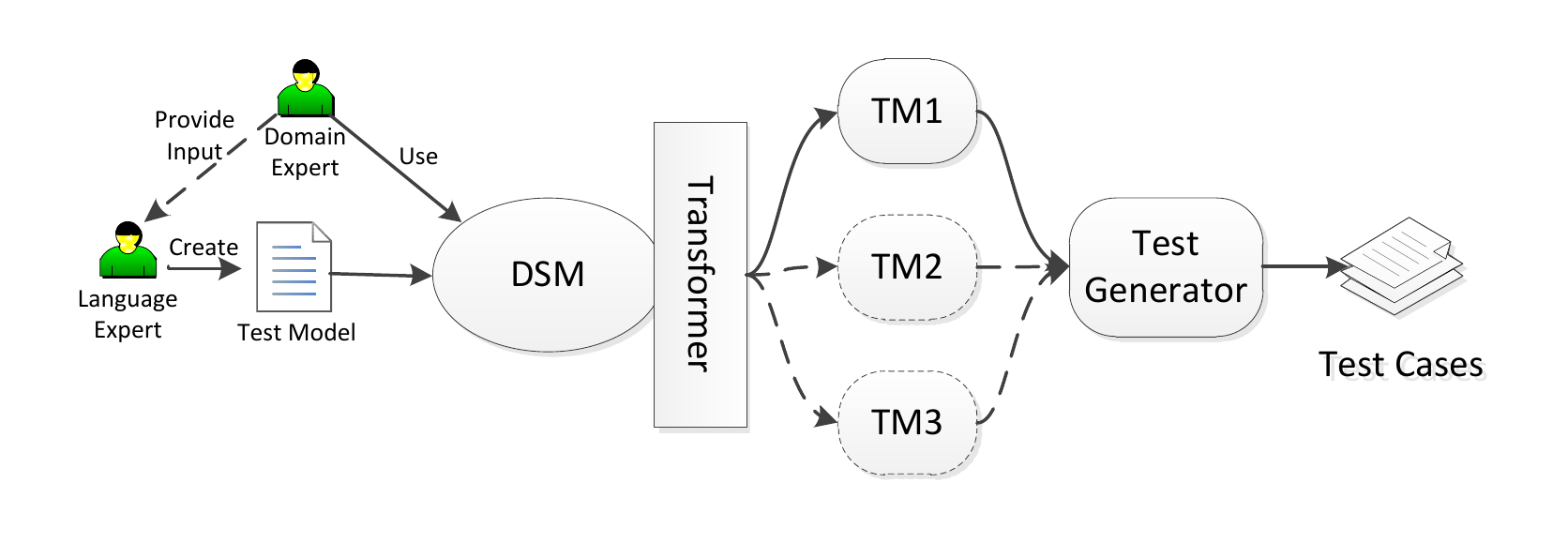} 
\caption{DSM elements in our approach}
\label{fg:dsm-elements}
\end{center}
\end{figure}

From the test automation perspective, our approach in terms of DSM is illustrated in Figure \ref{fg:dsm-elements}. Together with the domain expert, the language expert defines the generic overall test model. From this model, OSMOTester automatically forms a higher-level DSM language that the domain expert can use to guide the test generation from this model. In DSM terminology, a transformation is applied by OSMOTester based on the constraints defined by the domain expert to produce constrained variants of the test model (TM1, TM2, TM3 in the figure). The OSMOTester test generator then generates test cases from these test model variants (or scenarios/slices as in \cite{grieskamp2011}).

\subsection{Language and Generator Development}

The modeling notation supported by OSMOTester is the standard Java programming language with specific annotations and modeling objects to support test generation from the models. With the term model object we refer to Java classes containing transition methods and data-flow variable usage. The basics of the modeling notation and the composition of test models from this notation have been described in our previous work \cite{teemu2011}. It is based on our experiences in test modeling and use of other tools (e.g., ModelJUnit \cite{utting2007} is a close relative). We recall here the basics of this modeling notation including the latest evolutions.

We view the models for model-based testing as composed of two main elements: control-flow and data-flow. We start here by presenting the control-flow aspects followed by the data-flow aspects.

\subsubsection{Control-Flow Modeling}
\label{sec:control-flow}

With control-flow we refer to the order in which the different test steps are executed. The test step refers to executing a transition method on the model program. As is common for this type of approach, the possible transitions that can be taken at any time are defined by their associated guard statements. A transition can only be taken when the associated guard statements allow it to be taken.

To support for control-flow modeling OSMOTester defines two basic annotations, @Guard and @Transition. When a method in a model program is associated with a @Guard annotation, these methods must return a Boolean value of true to allow any associated transitions to be taken by the test generator. A guard method is associated to transitions by their naming, which is given as @Guard("name") and @Transition("name"). A guard can be associated to several transitions by using an array of names, or by leaving the name out completely which associates it to all transitions in the test model(s).

A transition is the central concept of the control-flow modeling as all other elements related to control-flow are executed together with associated transitions. As it is out experience that transitions in test generation represent a form of test step in a test model, we also support using the annotation @TestStep instead of @Transition. Other control-flow related elements include the @Pre and @Post annotations, which cause any tagged methods to be executed before (@Pre) or after (@Post) the associated transitions. Association is  similar as for @Guard annotations. For example, we have used @Post annotations to provide generic test oracles that should be executed after each test step (to evaluate model state vs SUT), and both @Pre and @Post annotations to provide logging information on the model state before and after test steps.

As described in \cite{teemu2011}, these elements can be embedded in different Java classes and then composed in a modular fashion by the OSMOTester tool. This allows one to specify different concepts, partial models, and their compositions in separate modules and configurations. Listing \ref{lst:cf-example} shows an example of a model object using these elements.

\begin{lstlisting}[caption=Model object example.,
  label=lst:cf-example,
  float=t]
/** The global model state, shared across test models. */
private final ModelState state;
/** The scripter for creating/executing the test cases. */
private final CalendarScripter scripter;

@Transition("AddEvent")
public void addEvent() {
  String uid = state.randomUID();
  Date start = state.randomStartTime();
  Date end = calculateEndTime(start);
  ModelEvent event = state.createEvent(uid, start, end);
  scripter.addEvent(event);
}

@Guard("RemoveOrganizerEvent")
public boolean guardRemoveOrganizerEvent() {
  return state.hasEvents();
}

@Transition("RemoveOrganizerEvent")
public void removeOrganizerEvent() {
  ModelEvent event = state.getAndRemoveOrganizerEvent();
  scripter.removeEvent(event.getUid(), event);
}
\end{lstlisting}

\subsection{Data-Flow Modeling}

As an extension to our previous work, we also provide a set of objects to support data-flow modeling. The main modeling objects for data-flow currently are:

\begin{itemize}\addtolength{\itemsep}{-0.5\baselineskip}
\item Value range: defining a numerical range of values for a variable
\item Value set: defining a set of values for a variable
\item Readable words: defining character combinations that consist of commonly displayed human-readable characters.
\end{itemize}

In order to use these data-flow objects, one initializes them at model-creation time and uses them to provide matching values where needed as input in the test steps. As each of the provided objects keeps a history of the values it has provided, it is possible to configure them with different algorithms for input generation. This includes random values, least covered values, or boundary values (for a value range). Each data-flow object can be configured separately with their own configuration of values and algorithms.

Listing \ref{lst:dataflow} gives an example of defining a range of possible values for a start date of a calendar event using a value range object. New values for this variable are generated using the next() method, which is part of an interface shared by all data-flow objects.

\begin{lstlisting}[caption=Data-flow example,
  label=lst:dataflow,
  float=t]
@Variable
/** Used to generate start times between January 2000 and December 2010. */
private ValueRange<Long> startTime;

public ModelState() {
  Calendar start = Calendar.getInstance();
  start.setTime(new Date(0));
  start.set(2000, 0, 1, 0, 0, 0);
  Calendar end = Calendar.getInstance();
  end.setTime(new Date(0));
  end.set(2010, 11, 31, 23, 59, 59);
  long startMillis = start.getTimeInMillis();
  long endMillis = end.getTimeInMillis();
  startTime = new ValueRange<Long>(startMillis, endMillis);
}

public Date randomStartTime() {
  return new Date(startTime.next());
}
\end{lstlisting}

Beyond these specific object types, also primitive values (e.g., integers, Strings, Booleans) and custom objects of any type can be recorded by OSMOTester. While their generation and updates may not be handled by OSMOTester itself, it will still record values of all @Variable tagged variables in the model objects regardless of their type. All values are recorded by their object instance, and can be compared to their String representation if used as definitions in coverage algorithms or similar components.

\subsection{Composing the Models and Generating Tests}

Once the different elements of the control-flow and data-flow models have been specified, they still need to be composed together and we need to be able to invoke a test generator to generate tests based on these models. OSMOTester provides the test generator component that basically traverses the given model program steps according to their guards and chosen generation algorithms. As the generator traverses the transitions, it also explores the data-flow space by generating data from any encountered data-flow object.

The binding of the model objects together in practice consists of creating the test generator (Java) object and using its methods to add the (Java) model objects to it. For space reasons, we do not show this as an example here but one is available in \cite{teemu2011}. The OSMOTester generator starts by parsing all the given model objects and associating all model elements to each other. It also stores references to all @Variable tagged variables to capture any values they produce. This allows for creation of more advanced algorithms and feature to support test generation without requiring specific action by the user.

It is possible to define a set of specific algorithms and constraints to guide the test generation also at this level. Test and suite end conditions provided with OSMOTester include:

\begin{itemize}\addtolength{\itemsep}{-0.5\baselineskip}
\item Length: end after generating a given number of steps or tests
\item Probability: end after any step or test with a given probability
\item Requirements coverage: end when the given requirements have been covered (identified by specific object calls in the test models)
\item Step coverage: end when the defined set of transitions has been covered
\item Data coverage: end when the defined set of values for given data variables has been covered
\item And/Or compositions: allow composing several end conditions together with logical operators
\end{itemize}

Similarly, different algorithms for traversing the given models can be defined based on the different elements of the model objects. OSMOTester includes the following algorithms:

\begin{itemize}\addtolength{\itemsep}{-0.5\baselineskip}
\item Random: picks a random transition from the ones available
\item Balancing: randomly picks an available transition but favors less covered ones
\item Weighted: randomly picks an available transition but gives a higher probability to ones with higher weight. 
\end{itemize}

Weight can be defined by the modeler as in @Transition("name", 5), where 5 is the weight.

\subsection{Domain Specific Scripting Languages}

In the previous subsection we described the first part of our DSM support, the language development framework. In this subsection we describe the second part of DSM in our MBT context: support for solution domain modeling. In our previous work we have created graphical domain-specific languages for manual test creation \cite{op2011} and for test modeling for model-based test generation \cite{teemu2012}. There approaches were based on using a specific domain-specific modeling tool to build the models separately from the test models.

Here the test models themselves are used as a basis for the domain language and the MBT tool as the test generator. The support for using such models has been integrated into the OSMOTester tool itself, requiring no external tools or practically no added effort on top of building the test models themselves as described in the previous subsection.

The basis for this domain language is formed by the transitions defined in the test model as described in section \ref{sec:control-flow}. More specifically, the names given to the transitions in the annotations as in @Transition("name") form the vocabulary of the domain specific language. In our experience, as people work to build such models and give names to their model elements, they naturally tend to use names of familiar domain and test concepts. For example, OSMOTester comes with a calendar application example that has the following transitions:

\begin{itemize}\addtolength{\itemsep}{-0.5\baselineskip}
\item Add event: Adds a new event to the calendar for an existing user.
\item Link event to user: Links an existing event to another user, making the new user a participant in the event organized by the first user.
\item Remove organizer event: Removes an event completely by deleting it from the organizer and as a result from all linked participants.
\item Remove participant event: Removes a participant from an event.
\item Add task: Adds a task for an existing user. A task is like an event with no duration or participants.
\item Remove task: Removes a chosen task from associated user.
\item Add task overlapping event: Adds a task that overlaps a chosen event in time for the same user.
\item Add overlapping event: Adds an event overlapping another event in time for the same user.
\end{itemize}

These are transitions (test steps) of the calendar example that describe its nominal (expected) behavior. Additionally, the calendar test model also defines a set of transitions for testing the error handling behavior of the calendar. These are the following transitions:

\begin{itemize}\addtolength{\itemsep}{-0.5\baselineskip}
\item Remove a task that does not exist: Tries to remove a task which does not exist (invalid data).
\item Remove an event that does not exist: Tries to remove an event which does not exist (invalid data).
\end{itemize}

OSMOTester supports two different types of applications of these elements as a domain-specific testing language. In this subsection we will further demonstrate their use in terms of the calendar example.

\subsubsection{Abstract Domain-Specific Scripting}
\label{sec:dsm-abstraction}

We can define constraints over the model programs to guide the OSMOTester test generator. The model defines the overall possible flows of test generation, while the scripting language allows for guiding it towards specific goals. This support is based on the observed needs in industrial application, where the domain experts wish to guide test generation to explore specific areas of interest at different times.

For example, considering the calendar example, we may wish to generate a set of test cases for having several overlapping events and some tasks, with several events active at a time (i.e., not removed before new ones are added). In this case, we can use the generic model program specified previously, and just define the specific requirements for the test cases in the scripting language. A script supporting this definition is shown in Listing \ref{lst:dsm-script}.

\begin{lstlisting}[caption=DSM script example.,
  label=lst:dsm-script,
  float=t]
setting, value
model factory, osmo.tester.examples.gui.TestModelFactory
algorithm, random

step, times
add event, >=2
add overlapping event, >= 3
add task, == 1
add overlapping task, <=2

variable, coverage
event count, 5
\end{lstlisting}

This shows three different tables illustrating different aspects than can be defined for configuring the test generator. First, we provide a settings table specifying that the model objects for test generation are provided by a class called osmo.tester.example.gui.TestModelFactory. We also define that we wish to use the random algorithm for test generation.

Second, we define a step coverage requirement table specifying that we want the AddEvent step (transition) to occur in each generated test case a minimum of 2 times. We also define that we want the step AddOverlappingEvent to occur a minimum of 3 times. The AddTask step needs to be present exactly once in each generated test case, and the AddOverlappingTask at most 2 times.

The third and final table in this example specifies that the variable EventCount should have the value of 5 at some point in each generated test case. As the model program uses the EventCount variable to express the total number of both organizer events and participant events, this ensures that their total sum as active events during a test case reaches 5 at some point. At the same time, the previously defined requirements for different types of transitions ensure that different types of events are also generated in each test case (organizer and participant).

A table missing in this example is one that allows the user also to override what data values are generated for specific input values. This table is similar to the variable coverage table, defining variable names and their possible values to be generated. If there is no definition for a variable, the generic model program definitions are used instead.

These examples also illustrate the expertise required to build suitable models than can be used in a diverse way as a basis for test generation. For example, in the calendar example, the current model program does not let us specify how many users we wish to have included in the generated tests. In fact, the example provided with the OSMOTester distribution generates a number of users randomly between 3 and 5. If this becomes important to control, a specific transition for adding transitions could be defined and we could then define exact number of these transitions required. Alternatively, a data variable could be defined that is used by the model program to generate a matching number of users and thus allows the user to control the number through the scripting language.

\begin{figure}
\begin{center}
\includegraphics[height=3.55in]{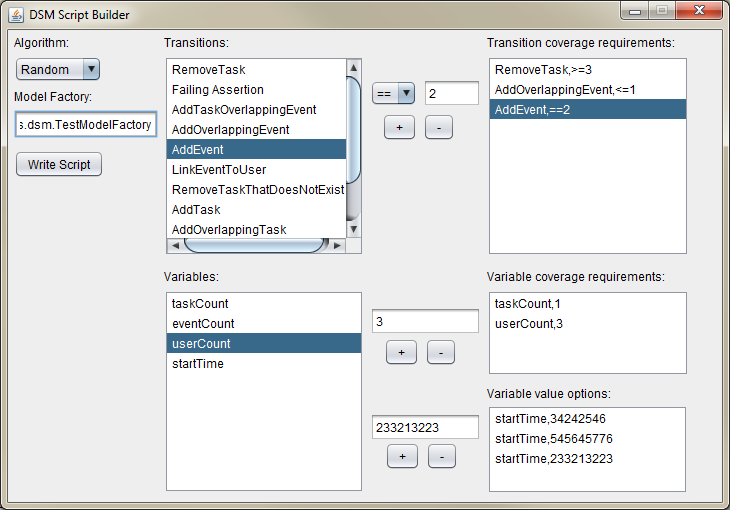}
\caption{OSMO DSM GUI} 
\label{fg:dsmgui}
\end{center}
\end{figure}

In addition to manual script writing, OSMOTester also provides a graphical user interface (GUI) to create these scripts. This is illustrated in Figure~\ref{fg:dsmgui}. The useful part of this GUI is that given the set of model objects that compose the model-program it can automatically identify all possible transitions and data variables. Thus it can show the user the options available for building the DSM scripts and also the operators available. An example of these is shown in Figure~\ref{fg:dsmgui} for the calendar example.

Considering the end result, generating a number of tests using test generation algorithms to execute a model program produces a lot of variation but this may not always be the type of variation the domain expert wants at that time. Many MBT tools use static analysis techniques such as symbolic execution and constraint solving to generate specific test cases traversing the model in desired paths. As we take a more dynamic approach of executing the model as a normal program, we also provide support for optimizing the test set in this way.We support optimizing for transition, transition pair, requirements (specific definitions in model objects), variable (number of values) and variable value (specific values) coverage via a greedy optimizer. The optimizer takes a weight value for each of these coverage criteria and provides an optimized subset of specified size from the given overall set.

In our experience test generation from the models is generally fast, and this provides a practical method for providing an optimized test suite. For example, generating 50000 test cases from the calendar example takes about 11 seconds, and optimizing this to find an optimized set of 50 tests with the algorithm takes about 17 seconds on an Intel CoreI5 2,67Mhz system utilizing a single core, with the Java virtual machine maximum heap size of 880MB (formed by Java default settings, not modified or monitored as memory use did not become an issue). As the test object OSMOTester creates for each generated test case gives the model program a possibility to store the test script with it, the user can then pick the optimized set of scripts for use.

%\begin{figure}
%\begin{center}
%\includegraphics[height=3.05in]{50-raw.png} 
%\caption{Unoptimized set of 50 generated tests}
%\label{fg:50raw}
%\end{center}
%\end{figure}

%\begin{figure}
%\begin{center}
%\includegraphics[height=3.05in]{50-long.png} 
%\caption{Optimized set with length weight 20}
%\label{fg:50long}
%\end{center}
%\end{figure}

%\begin{figure}
%\begin{center}
%\includegraphics[height=3.05in]{50-short.png} 
%\caption{Optimized set with length weight -20}
%\label{fg:50short}
%\end{center}
%\end{figure}

\subsubsection{Manual Domain-Specific Scripting}

Model-based testing is commonly considered as a means for automated test generation from test models. The tests are generated by using some automated algorithm to traverse the test model as illustrated in previous sections. While it is a great goal to strive for creating all test cases automatically from such models, in practice it is our experience that many people still wish to see a set of specific manually crafted test cases that they can control and verify they cover a given set of paths, requirements and other elements of interest. The approach we presented in section \ref{sec:dsm-abstraction} helps address many of these requirements through guiding the test generation through user-provided constraints. However, additionally even more explicit options for manual guidance are needed to fully address this need.

\begin{figure}
\begin{center}
\includegraphics[height=3.55in]{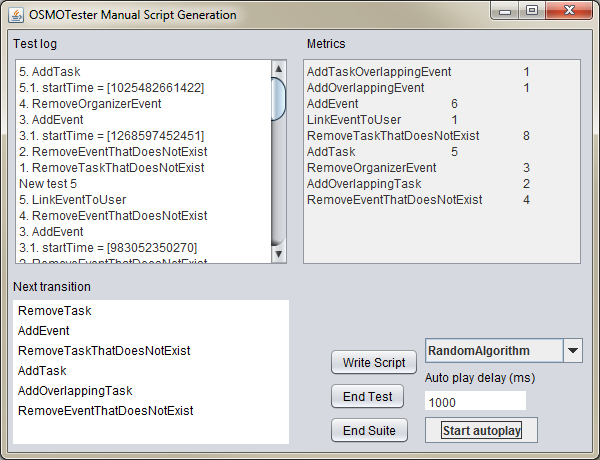} 
\caption{Manual GUI}
\label{fg:manualgui}
\end{center}
\end{figure}

In this section we present how OSMOTester enables the user to manually generate exact test cases from the same test models. Similar to the DSM scripting language, this can also be manually scripted or guided via a GUI shown in Figure~\ref{fg:manualgui}. This GUI consists of four main elements. The test log in the upper left corner shows the test steps that have been taken. The sequence number is reset when a new test case is started. The metrics set in the upper right corner shows how many times the different transitions have been taken in the current test suite. 

The next transition choice in the lower left corner shows which transitions can be taken at a given time. By clicking on a transition in this list, the user can guide the OSMOTester generator to take that transition. Choosing one actually executes that step of the model program and updates the set of available transitions. As the GUI only displays the valid transitions for the current state, the user can only create valid test cases with it. 

The lower right corner contains the set of controls used to further guide the test generation. The write script button writes the generated test script to a file. It should be noted that this is not a test script for the SUT but a script for OSMOTester itself, defining which transitions to take and which values to provide for data variables. The end test button starts a new test case for the suite and end suite finishes the test generation. The algorithm, delay and start autoplay controls are all related to the autoplay feature. It enables the user to start and stop automated model traversal using a chosen algorithm, and to continue manual generation when preferred, always updating the GUI.

As an end result, OSMOTester will generate a test script matching the set of test cases and their transitions (test steps) as generated in this tool. Listing \ref{lst:manual-example} shows an example portion of this script matching the end of test 4 and start of test 5. Notice that this also includes the values of model variables used in those steps. That is, the user can script both the sequence of transitions and variable values in those steps.

\begin{figure}
\begin{center}
\includegraphics[height=1.05in]{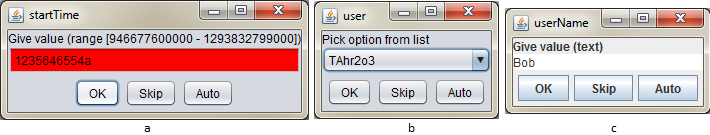} 
\caption{Dataflow GUI (a) Value Range (b) Value Set (c) Readable words}
\label{fg:dataflow}
\end{center}
\end{figure}

Beyond the transitions, the GUI also allows the user to specify the data values of relevant model variables. Figure\ref{fg:dataflow} shows an example of three types of dataflow variables. The (a) option shows a request for the user to provide a value for a numeric range (shown in red if an invalid number is give), (b) shows a choice from a set of values, and (c) shows a request for a word that can be any set of readable characters. The name of the variable is shown in the window title. Option OK inputs the given manual choice for this step in the model program, Skip uses the model program logic to generate the value once, and Auto sets the model program to generate all values for this variable in the future automatically.

\begin{lstlisting}[caption=Manual test script example,
  label=lst:manual-example,
  float=t]
action, name, value
...
step, link event to user,
new test,,
step, add event,
variable, start time, 1268597452451
step, remove event that does not exist,
step, remove task that does not exist
\end{lstlisting}

This approach has several benefits. First of all, it enables the user to easily create specific test cases from the same model as is used for test generation. Thus test scripting can be done from a single source with less duplicated effort. It also eases test maintenance as updates to the model will also be directly updated to the test cases themselves. Thus changes required to test cases to adapt to changes in the test target can be handled in one place and are automatically updated to all test cases, both manually created and automatically generated.

\section{Discussion}
\label{sec:discussion}

As we have presented, our approach is based on using existing programming languages and tool as a basis for modeling. As the approach is only based on the basic programming language constructs, it does not require specific tool support for modeling such as a customized visualization and editing component. Instead, the user can choose any integrated development environment (IDE) they are familiar with and use that as they find best (within the Java programming language constraints). This also makes it much easier for us to build a modeling language and framework as we do not need to worry about features such as refactoring, syntax highlighting, debugging and code navigation. These are practically provided for free by the available tools (IDE). 

In our experience, people working with test automation are often also experienced programmers. For them, working with this type of a modeling approach is easy as they can work using tools and notation familiar to them. This also keeps them informed exactly about how their tests are built, instead of hiding if behind the custom notations of tools. In practice our experience is also that the approach gives a lot of power that is not available in customized notations as popular general purpose programming languages typically have more resources put into their development and as well as available libraries.

However, as not all people are experts in programming or wish to write their test case using a programming language notation, higher levels of abstraction are also important. For example, a domain expert may be interested in testing their system using specific configurations of the test models with specific properties. Even if in theory it may be possible to show that a MBT tool will generate a set of impressive test cases, it is equally or even more important to provide confidence for different stakeholders that the important features and their important properties are covered in the provided test cases. By providing means to generate overall test cases from the model, constrain the test generation by using a DSM language, or to manually define specific test cases we enable these different goals to be achieved.

Many existing tools make extensive use of static analysis techniques such as symbolic execution and constraint solving \cite{utting2007}. These enable features such as generating test data to cover specific paths of the test model. In our case, we have opted not to use such techniques, beyond those provided by the IDE's, but to focus on supporting and effectively exploiting the runtime execution aspects. In this case, any programming language features can be used  in the model programs without worrying about the scalability of the analysis techniques. At the same time, as the model program is executed it can also be easily used to provide features such as the manual test generation GUI we have presented here. As the model program is actually executed, the user can at all times be given exact information about the current state of the model and what actions are possible in that state.

While the approach in many ways gives the user power, it also puts a lot of responsibility to them. As the models are not extensively checked by the tool, they may require some more analysis effort by the user (which is not necessarily a bad thing as it increases the understanding of the system). For example, one has to realize that requiring 10 event removals will never terminate if only 5 event creations are allowed. The domain expert needs to have such understanding, but it is also possible to configure end conditions and create models that report such failures, or provide thresholds for breaking and report failure to achieve required constraints, if needed.

For optimization we apply test selection by allowing one to generate a large set and provide optimization algorithms to select a subset with specific properties. This works when generating offline test scripts to be executed after generation. However, if done in online mode where the tests are executed as generated this is not effective but rather manual tuning of the model becomes more important. In these scenarios it could be beneficial also to have more advanced analysis features from the static analysis domain as well, such as the model checking capabilities in Spec Explorer that also uses a form of model programs \cite{grieskamp2011}.

OSMOTester is currently being used by several companies in the embedded and software industry to support test automation. It has been our experience in case studies with industrial partners that the user naturally names the elements in the built model using familiar terms in the target domain. It has also been our experience that while a language expert (a role we commonly take) may help them build a model, they can make the most of it if given a simple way to guide the test generation for different aspects in their model. Yet in many cases they are not interested in learning all the details of the model internal elements as domain experts. Examples of real requirements include setting up a model to produce a certain number of users, varying the generated tests with a chosen subset of possible parameters, and using a specific set of values in specific set of generated tests. The approach we have presented here has helped address those needs.

\section{Conclusions}
\label{sec:conclusions}

Model-based testing is a powerful test generation technique and tools for this have come a long way and are increasingly adopted in the industry. However, their generic nature, required skills and exposed formalisms can hinder their potential in industrial context with domain experts. In this paper, we presented an approach to guide model-based test generation using domain-specific concepts through support integrated into a MBT tool. This approach starts from using a framework over a common programming language (Java) to build the test model program. The naming conventions used in this framework automatically turn this test model into a domain-specific language that can be used to guide the test generation from this model. The user can then either generate test from the generic test model, constrain it to generate varied test cases for specific scenario(s) or manually create specific test cases from the model.

This allows not only automated generation of a large set of test cases from the test model but also addressing more specific needs by domain experts without needing a language (test model) expert to help with model customization. It also allows one to build more confidence in covering specific elements of the system under test as required by test requirements. As the user can create test cases from the model using manual guidance they can have confidence the requirements are covered and how they are covered. This also helps in the general test maintenance issues as updating the model will also automatically update the manually created test scripts that are based on this model. In the future we plan to explore further means to ease the adoption and use of MBT with concepts such as specification mining input.

\nocite{*}
\bibliographystyle{eptcs}
\bibliography{references}
\end{document}